# Proof-of-principle experiment for nanoparticle-assisted laser wakefield electron acceleration

Constantin Aniculaesei[1], Vishwa Bandhu Pathak[1], Kyung Hwan Oh[1], Prashant Kumar Singh[1], Bo Ram Lee[1], Calin Ioan Hojbota[1,3], Tae Gyu Pak[1,3], Enrico Brunetti[4], Byung Ju Yoo[1], Jae Hee Sung[1,2], Seong Ku Lee[1,2], Hyung Taek Kim[1,2,#], Chang Hee Nam[1,3]

[1]Center for Relativistic Laser Science, Institute for Basic Science (IBS), Gwangju 61005, Republic of Korea.
[2]Advanced Photonics Research Institute, Gwangju Institute of Science and Technology (GIST), Gwangju 61005, Republic of Korea
[3]Department of Physics and Photon Science, GIST, Gwangju 61005, Republic of Korea.
[4]Scottish Universities Physics Alliance, University of Strathclyde, Department of Physics, Glasgow, G4 0NG, United Kingdom.
#corresponding author: htkim@gist.ac.kr

We demonstrate for the first time a proof-of-principle experiment for nanoparticle-assisted laser wakefield acceleration. Nanoparticles generated through laser ablation of an aluminium target were introduced into a helium plasma and used to trigger the injection of electrons into the nonlinear plasma wake excited by an 800 nm wavelength, 1.8 J energy, femtosecond duration pulse laser. High-energy electron beams were produced, observing a significant enhancement of the electron beam energy, energy spread and divergence compared with the case when electrons are self-injected. For instance, the best quality electron bunches presented peak energy up to 338 MeV with a relative energy spread of 4.7% and a vertical divergence of 5.9 mrad. The initial results are very promising and motivate further theoretical and experimental research into developing the nanoparticle-assisted laser wakefield acceleration.

## INTRODUCTION

A laser wakefield accelerator[1] (LWFA) uses the plasma waves generated from the interaction between a high power laser and a plasma to accelerate electrons to relativistic energies. In the highly nonlinear case of laser-plasma interaction called the "bubble"[2] or "blowout"[3] regime, an intense laser pulse pushes plasma electrons forward and sideways leaving behind a spherical region ("bubble") filled with ions. The displaced electrons form a sheath around the bubble creating an acceleration wakefield that can exceed 100 GV/m. Some of the electrons that cross at the base of the bubble can be injected into the bubble in a process called wave-breaking[4,5]. The condition for electron self-injection is that the electrons group velocity is greater than the phase velocity of the back of the bubble[6]. The electron injection carries on until the space-charge effect of the injected electrons is strong enough to stop further injection. Due to the highly nonlinear evolution of the plasma waves, the onset of self-injection depends on the initial conditions of the laser pulse and plasma medium thus leading to some degree of randomness in the self-injection process, consequently affecting the parameters of the accelerated electron bunch. To overcome the randomness of the injection process various schemes have been developed to manipulate the longitudinal momentum of the electrons surrounding the plasma wake[7,8] or the phase velocity of the plasma wake[9,10].

There have been theoretical studies on the use of a nanowire or nanoparticle for seeding electrons in LWFA. The use of a nanowire in the bubble regime was proposed by Shen et al.[11] in 2007 and the use of a nanoparticle was investigated theoretically in 2018 by Cho et al.[12] with multi-dimensional particle-in-cell (PIC) simulations. These new electron acceleration schemes use a nanowire or a nanoparticle to trigger the injection of electrons in the acceleration phase of the plasma wakes. It was shown that a nanoparticle/nanowire ionized by the leading edge of the laser pulse, creates a strong electric field that attracts some of the plasma electrons, which can gain additional momentum and be injected into the bubble. These theoretical studies predicted that the characteristics of the accelerated beam depend on the nanoparticle composition, size or number and also on the moment/position when the nanoparticle is injected into the bubble. The composition or the size of the nanoparticle controls the number of electrons injected in the wake, and therefore the charge of the accelerated bunch. The number of nanoparticles controls the number of electron bunches injected into the wake thus controlling the bunch length, structure and total charge. Injecting the electrons at earlier or later times leads to accurate control of the electron beam energy by increasing the acceleration length and ensuring that the injection can happen when the strength of the accelerating field of the wake is the highest. Thus, the nanowire/nanoparticle-assisted wakefield acceleration (NA-LWFA) could be a versatile scheme due to its high degree of control over the wakefield acceleration process.

Since this kind of scheme was first proposed more than 12 years ago, there have been, to the author's best knowledge, no experimental attempts to test the viability of NA-LWFA. For best performance, the NA-LWFA scheme would require an aerodynamic lens system[13] (ALS) that delivers a stream of nanoparticles that can be injected[12], ideally one nanoparticle at the time, into the plasma wake with micrometre precision. An ALS is not only bulky and expensive but also requires a serious engineering effort to integrate it with a gas target.

We decided to investigate first the feasibility of a simplified NA-LWFA scheme without employing an ALS device. As a source of nanoparticles, we used laser ablation[14] and we integrated it with a supersonic gas jet. This alternative scheme lacks the possibility of controlling the position where the nanoparticle is injected in the plasma wake, but the nanoparticle size and number can be controlled[15] by tuning the fluence of the laser beam used for ablating material. The nanoparticle + helium gas mixture then supersonically expands in a vacuum chamber, interacting with a high power femtosecond laser and generating an electron beam. In the present case, the ablation laser fluence is reduced until the density of nanoparticles is, theoretically, low enough to assume that the main laser interacts with only a few nanoparticles. The reduction of the laser fluence has been done while monitoring the electron beam parameters as shown in Methods. The results obtained from this proof-of-principle NA-LWFA experiment are presented in the next section followed by a brief discussion regarding some aspects of the electrons injection in the plasma wake in the presence of a nanoparticle.

## RESULTS

The experiment has been run at the Center for Relativistic Laser Science (CoReLS and the setup is shown in Figure 1.The setup installed in a vacuum chamber maintained at a pressure of $10^{-5}$ mbar before opening the gas valve consisted of a high power laser, focusing optics, gas nozzle and electron beam diagnostics. The setup is further described in detail in Methods. The accelerated electron beam has been characterized in terms of vertical and horizontal divergence, peak energy and relative energy spread. The peak energy is defined as the electron energy where the spectrum presents the highest charge density. All the shots are obtained for a helium pressure of 10.5 bar at the nozzle inlet, corresponding to a measured plasma density of $5 \times 10^{18}$ cm$^{-3}$ at a distance of 1.5 mm above the nozzle. The main laser was focused at the beginning of the gas jet density ramp at a fixed height of 1.5 mm above the nozzle. These settings have been chosen after a systematic scanning of the experimental parameters until the most stable electron beam has been obtained with helium gas. When used, the ablation laser pulse arrives 10 microseconds before the main pulse, a delay empirically set after systematic scanning until the best quality electron beam has been obtained with nanoparticle-assisted electron injection.

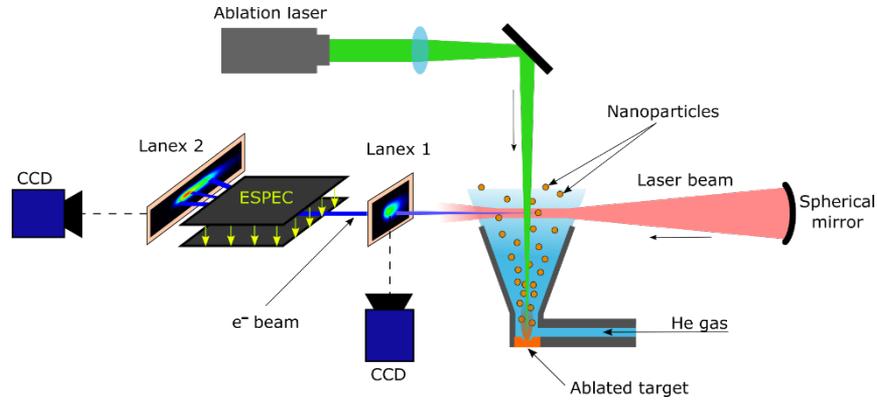

*Figure 1: The setup used in the nanoparticle-assisted laser wakefield acceleration. The nanoparticles generated through laser ablation mix with helium gas and co-propagate towards the outlet region where the high power laser excites nonlinear Langmuir plasma waves in the bubble or blowout regime. Accelerated electron bunches are diagnosed using Lanex1 (transverse profile, pointing, divergence and charge) and Espec+Lanex2 (energy and energy spread) (Espec: electron spectrometer, CCD: charge-coupled device).*

We recorded two data sets: with and without nanoparticles present in the helium gas jet. From each dataset, we selected for the present work the best (smallest energy spread) 10 shots, as displayed in Figure 2 and Figure 3. A summary of the results is shown in Table 1 in Supplementary Material. In the case of LWFA with electron self-injection, the quality of the electron beam was very poor, with relative energy spread (shown in Figure 2a) larger than 30% and divergence (shown in Figure 3a) larger than 12 mrad. A significant enhancement in electron beam quality was obtained when nanoparticles triggered the electron injection. In this case, the relative energy spread (shown in Figure 2b) was as small as 4.5% while the divergence (shown in Figure 2b) decreases to 6 mrad. Most of the shots recorded with nanoparticles shared the same specific signature: low divergence, electron bunches very well defined spatially and in energy, with a very low energy spread and higher peak energy. Switching on and off the nanoparticle source alternatively it was easy to observe that the electron beam quality changed also depending on the presence of the nanoparticles. A small number of shots (about 10% of the total number of 74 shots) had very poor quality, which could be due to the main laser pulse missing the nanoparticles. This was possible because the number of nanoparticles, according to theoretical calculations (see Methods), was very small.

It is interesting to observe that the peak electron energy recorded when nanoparticles are present is consistently higher than what is obtained with self-injection. This result could be due to two reasons: space charge effects and acceleration length. As shown in Table 1 (Supplementary Material) the charge is much smaller than what is obtained with self-injection. Since the charge loaded into the bubble modifies locally the acceleration field of the plasma wake smaller charge results in a higher effective acceleration field. Another factor that determines the final energy is the distance over which the electrons are accelerated (the acceleration length). Three-dimensional PIC simulations (not shown here) using the real experimental parameters indicate that electron self-injection happens almost at the end of the gas jet. In contrast, in the case of NA-LWFA, the electron injection can happen as soon as the plasma wake encounters an ionized nanoparticle. Since the nanoparticles are assumed uniformly distributed in the volume of the gas jet it is very probable that the nanoparticle-triggered electron injection could happen at much earlier times than the self-injection leading to a much longer acceleration length.

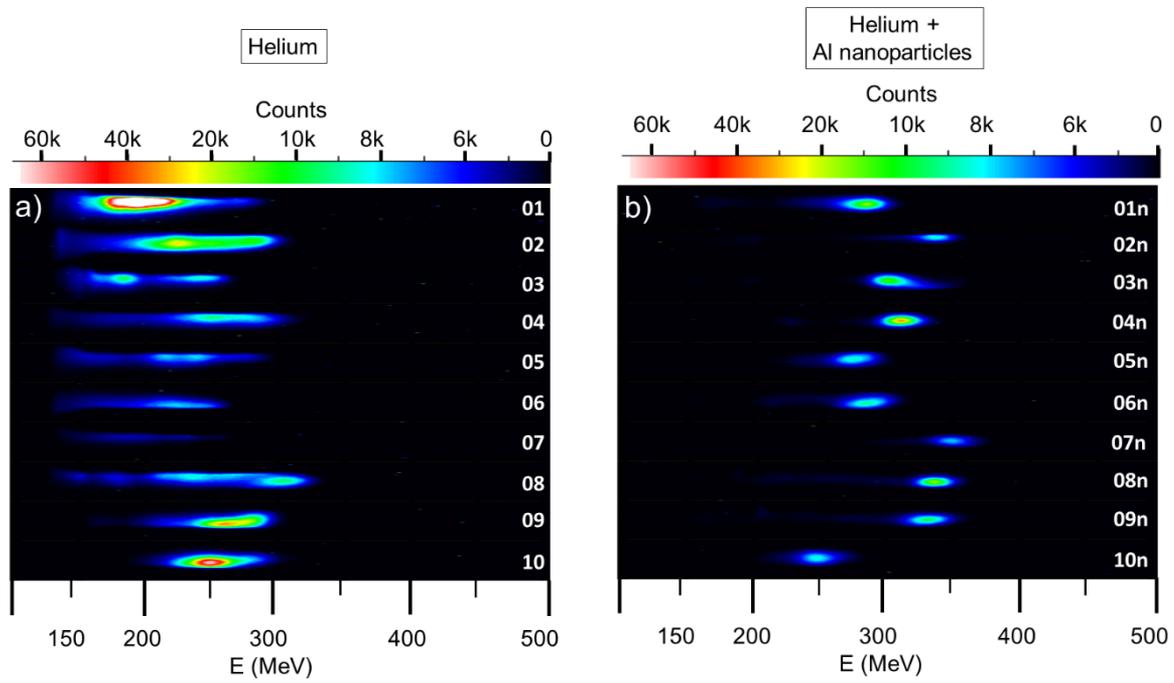

*Figure 2 False-color images of the electron energy spectra recorded in the electron spectrometer. a) Electron spectra generated without assistance from nanoparticles and b) The best energy spectra generated with nanoparticles. When nanoparticles are used to control the electron injection in the "bubble" the electron energy spread is significantly reduced, as compared with the case without nanoparticles*

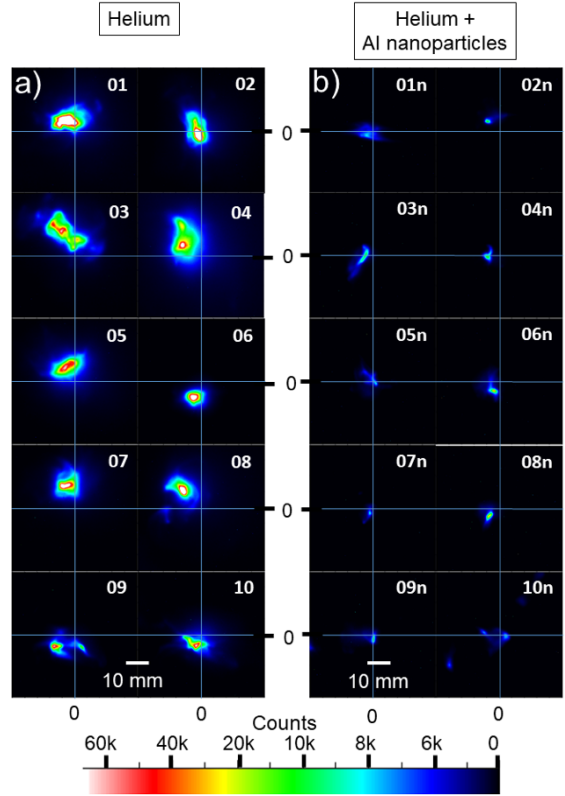

*Figure 3 False-color images of the electron beam transverse distribution recorded with Lanex 1. a) Electron beam profiles obtained without nanoparticles and b) Electron beam profiles obtained with nanoparticles. When nanoparticles are used the electron beam divergence is reduced, as compared with the case without nanoparticles.*

## DISCUSSION

To get an insight into the process of electron injection via NA-LWFA[11,12] we run two-dimensional particle-in-cell (PIC) simulations using OSIRIS [16]. Details about the simulations can be found in Methods. Figure 4 shows the time evolution of plasma density (a1 to f1) and of longitudinal momentum in phase space (a2 to f2) to clarify the role of nanoparticles in the injection. As the laser pulse propagates through the pre-ionized plasma, it pushes background plasma electrons (BPE) away from the axis, exciting a non-linear space-charge field, also referred as wakefield, just behind the laser pulse due to the immobile ions (Figure 4 a1). When the laser encounters a pre-ionized and charge-neutral nanoparticle, it pushes the nanoparticle electrons (NPE) away from the axis, and excites a localized space-charge field which grows stronger as the rising edge of the laser pulse passes through it (Figure 4 a2-c2), imparting extra momentum to the electrons in the vicinity of the laser pulse (Figure 4 c2). As the nanoparticle approaches the falling-intensity edge of the pulse, the strength of the ponderomotive force decreases. The BPE, as well as the fraction of NPE which were trapped in the front part of the laser, are trapped in the nanoparticle potential and travel with the nanoparticle as it passes through the nonlinear wakefield (Figure 4 d2). As the nanoparticle approaches the back of the wakefield (Figure 4 e2), the effective field around the nanoparticle again changes in the presence of the strong plasma sheath created by the crossing of BPE. This results in the release into the wakefield of the electrons trapped in the potential of the nanoparticle (Figure 4 e2 and f2). The interaction between the nanoparticle and the laser (Figure 4 b1-f1), also triggers transverse plasma modulations[11] at the front of the laser pulse. Plasma electrons in this region, due to snow-ploughing effect[17], stay in the front part of the laser on longer time scale. Therefore, the transverse modulations also stay with the laser, extending at the same time to the whole wakefield region and evolving transversely.

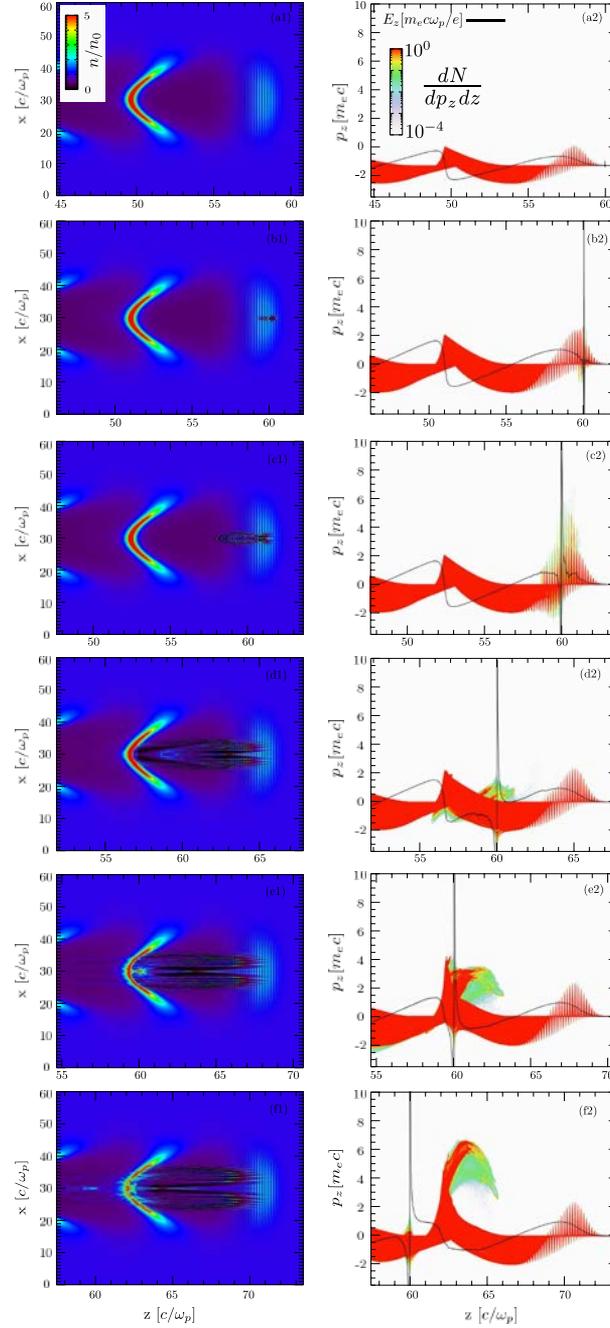

*Figure 4: 2D PIC simulations highlighting the nanoparticle-assisted trapping of plasma electrons in the non-linear wakefield excited by the laser. Snapshots of plasma density profiles at progressing times are shown in (a1) - (e1), and corresponding phase-space diagrams are shown in (a2)-(e2). The solid lines in (a2)-(e2) are the on-axis longitudinal fields, and the spike in the fields are due to the nanoparticle space charge. A fully-ionized nanoparticle is initialized at $z = 60 c/\omega_p$, and the window is moving with velocity c following the laser pulse.*

These density modulations can strongly diffract the laser pulse. This may explain why self-injection is not observed experimentally, in the presence of nanoparticles. It is also worth mentioning here that in our simulations we do not observe any injection caused by these transverse density modulations as reported by

Shen et al.[11]

The experimental results presented in this work show that the nanoparticle-assisted injection scheme for laser wakefield acceleration produced much better quality beams than the self-injection scheme. We want to emphasize, however, that we do not claim that self-injection always generates poor quality electron beams. Results in the literature show that good quality electron beams can be obtained using the self-injection scheme[18–21] to obtain, using the self-injection scheme but the exact laser parameters are not yet fully identified. In this experiment, self-injection was very hard to achieve and, overall, produced very poor quality electron beams. During the experiment, we measured strong nanosecond laser pre-pulses of the order of $10^{-4}$ relative to the main pulse, thus we can assume that the energy in the main pulse is much lower than what we measured using the full beam. The presence of the pre-pulse could also lead to Coulomb explosion of the nanoparticle[11] but this effect, if present, did not seem to affect the quality of most of the electron bunches. Even in these conditions, where the laser energy could be close or lower than required for self-injection, the nanoparticle-assisted electron injection significantly enhanced the electron beam quality as predicted theoretically[11]. This proof-of-principle experiment of NA-LWFA can be further improved using the same setup presented in this work but with a more stable laser. This should enhance the electron beam shot to shot stability thus permitting a systematic scan of more experimental configurations, such as the use of femtosecond ablation laser or different ablated materials.

The results presented in this work could have a strong impact on the development of femtosecond laser wakefield accelerators operating at kHz repletion rates[22]. In such accelerators, technology constraints currently limit the laser energy to a few 10's of mJ thus requiring high-density, small-length gas targets to produce an electron beam. Using nanoparticle-assisted electron injection could boost the electron energy and significantly improve the beam quality thus opening the road towards the development of a wide range of applications of the kHz, mJ laser wakefield accelerators.

The present work has proven that NA-LWFA is a viable scheme. It is, therefore, worthwhile to invest time, effort and money to integrate an ALS with a gas target allowing better control of the charge, energy, energy spread and bunch structure of the accelerated electron beam.

Our 2D PIC simulations indicate that the bunch length at the end of the acceleration process is of the order of 1 micrometre (or less than 3.3 femtoseconds long) suggesting that attoseconds bunches could be produced if the number of electrons injected can be better controlled. The exact conditions for the production of shorter electron bunches using a NA-LWFA scheme could be determined using 3D PIC simulations. If proven to be experimentally realizable, this technique could open new ways for the development and application of ultrashort radiation sources[23–25].

## MATERIALS AND METHODS

### The gas + nanoparticles target system

The gas jet + nanoparticles target consisted of two main parts: the supersonic nozzle and the source of nanoparticles. The supersonic nozzle was a conical *de Laval* type of nozzle with 2 mm inlet diameter, 3.9 mm outlet diameter and 8° semi-opening angle. The helium gas was fed into the nozzle through a 2 mm diameter pipe placed sideways (see Figure 3). The gas target was mounted on a motorized stage that controlled the movement of the nozzle in all three spatial directions. To limit the amount of gas released in the chamber during the experiment, the gas was fed through a solenoid valve (Parker Series 9) and operated in pulsed mode with an opening time of 6 ms; the laser pulse interacted with the gas jet 3 ms after valve opening. During the experiment, the gas jet profile has been continuously monitored using a Fresnel-type of the interferometer (not shown here) to ensure that the density profile has a uniform distribution in all the experimental conditions.

The bottom part of the nozzle is a removable 2 mm thick aluminium plate fixed with screws. The nanoparticle source is based on laser ablation. The ablation laser (AL) with a wavelength of 532 nm, 3 ns pulse length and a beam size of 5 mm is focused down into the gas nozzle and ablates the surface of the removable plate. Previously published results[14] have shown that the ablation plume produced by a nanosecond, picosecond or femtosecond laser can contain significant amounts of nanoparticles and micron-sized structures. The density and size of nanoparticles generated through laser ablation depend on the laser fluence, laser pulse length and the type of material ablated. The ablation plume mixes with the helium gas and is transported towards the outlet of the nozzle where it interacts with the main laser beam. The energy of the AL has been empirically set to 10 mJ per pulse, for which the divergence and energy spread of the electron beam was the lowest. In these conditions, the AL fluence on the target was 10.39 J/cm$^2$, which, according to published works[14,15], creates nanoparticles with sizes of the order of 10's of nm. This assumption has been confirmed by atomic force microscopy measurements (not shown here) which showed that the majority of generated nanoparticles had diameters between 10 nm and 20 nm. Optical probing using the interferometry could not show any phase shift when interferograms with and without nanoparticle present have been compared, which means that the nanoparticle density is much lower than the sensitivity limit of the interferometer. Using a theoretical estimation based on the amount of ablated material[15] and assuming that all the material was transformed into spherical nanoparticles with 20 nm diameter we obtained $10^8$ nanoparticles per shot. This estimation was based on idealized parameters, thus the number of nanoparticles is expected to be much lower. Changing the time delay between the ablation laser and the main laser and monitoring the quality of accelerated electron beam it was easy to observe that the nanoparticles influence can be seen for approximately 3 microseconds. Taking into account the gas jet speed, which was supersonic (1400 m/s), we can estimate that the nanoparticles were present in a cylindrical volume with 4 mm diameter and 4.2 mm height. The density of nanoparticles can be approximated then to be about $10^7$-$10^8$ nanoparticles/cm$^3$.

**The laser wakefield accelerator**

The Ti:Saphire high power laser, with a central wavelength of 800 nm, horizontal polarization, based on the chirped pulse amplification technique[26] delivered 1.8 J pulses (after compressor) with a temporal duration of 27 ± 2 femtoseconds at a repetition rate of 5 Hz. A spherical mirror with the focal length of 1 meter focused down the 65 mm diameter laser beam to a focal spot with a diameter of 40 microns measured at full-width-at-half-maximum (FWHM), which contained 60% of the total energy. This ensured on target a normalized FWHM vector potential $a_0 = 1.07$.
For the entire experiment the laser beam was focused at the start of the density ramp, 2 mm longitudinally from the centre of the nozzle, with the spot placed horizontally in the centre of the gas nozzle and vertically at a distance of 1.5 mm above the nozzle exit. The transverse and longitudinal alignment was done with a Top View imaging setup which consists of a CCD camera (FLIR BFLY-U3-50H5M-C) equipped with a camera lens (Samyang 135 mm F2.8 ED UMC) and a BG39 glass filter to cut the scattered light from the laser. The Top View setup was also used to monitor the optical emission from the plasma channel[27,28].

Lanex$^{TM}$ 1 setup consisted of a 6×6 cm Gd$_2$O$_2$S:Tb screen (Lanex$^{TM}$ Back) placed at 410 mm from the gas jet and imaged onto a CCD camera (FLIR BFLY-U3-50H5M-C) with a camera lens (Samyang 100 mm F2.8 ED UMC) and provides information about the transverse electron profile, pointing, divergence and relative charge. The results for each dataset (taken under the same experimental conditions) were averaged and the mean value and the mean standard error was calculated.

The electron spectrometer (E-spec) consisted of a 0.996 T dipole magnet that has an opening with a height, width and length of 8×70×205 mm. The entrance of the E-spec was placed at a distance of 455 mm from the gas nozzle and dispersed electrons in the horizontal direction as a function of their energy onto a 430×80 mm $Gd_2O_2S$:Tb screen (Lanex[TM] Back) placed at 1155 mm from the gas nozzle. The Lanex[TM] screen was imaged onto a CCD camera (FLIR GS3-U3-50S5M-C) with a lens (Samyang 16 mm F1.4 AS UMC). The error in energy reading of the electron spectrometer was determined by the electron beam size and pointing and varied depending on energy between (+6, -3.2) MeV for 175 MeV and (+35, -15.7) MeV for 400 MeV. The energy calibration of the electron spectrometer was done using the G4beamline code[29].
After energy calibration, the peak energy (the part of the energy spectrum that contains most of the charge) was determined from the electron energy with the maximum counts in the image.

**Particle-in-cell simulation**

We performed a set of 2D PIC simulations using OSIRIS[16] in order to illustrate the physics behind the nanoparticle assisted trapping of electrons in the non-linear wakefield. A two-dimensional simulation box (x-z), moving with velocity c, is considered with dimensions $20 \times 60 \ (c/\omega_p)^2$ divided into $20000 \times 60000$ cells. We consider a preformed neutral plasma of two species: plasma electrons, initialized with $4 \times 4$ particle per cell, and nanoparticle electrons, initialized with $100 \times 100$ particles per cell. For the simulation we, consider a laser with $a_0 = 3.0$, FWHM spot diameter $D_0 = 40 \mu m$, and FWHM pulse length $\tau_0 = 30 \ fs$, propagating in a pre-ionized plasma with density $5 \times 10^{18} cm^{-3}$. A pre-ionized, 20 nm in diameter and $100 n_{cr}$ in charge density, nano size plasma is initialized on the laser axis.

We could not run three dimensional (3D) PIC simulations due to the high grid resolution required to resolve the nanoparticle which made the computational cost prohibitive for us. Also, one can notice that a higher $a_0$ is used in the 2D simulations as compared to the experiments. This is done intentionally to match the position of self-injection (without nanoparticle) in 2D simulations with the full scale 3D simulations taking into account the same laser parameters as used in the experiments, ie., $a_0 = 1.0$, FWHM spot diameter $D_0 = 40 \mu m$, and FWHM pulse length $\tau_0 = 30 \ fs$. We observe that in both cases the self-injection occurs at a plasma length of 5 mm. For the 3D (2D) simulations, without the nanoparticle, we consider a moving box, propagating with velocity c, of dimensions $20 \times 60 \times 60 \ (c/\omega_p)^3$ ($20 \times 60 \ (c/\omega_p)^2$) divided into $2000 \times 600 \times 600$ ($2000 \times 600$) cells, and 2 (16) particles per cell for the BPE.

**Acknowledgements**
This work has been supported by the Institute for Basic Science of Korea under IBS-R012-D1.
H.T.K., J.H.S., and S.K.L also supported by GIST through the "Research on Advanced Optical Science and Technology" grant in 2017.
E.B. acknowledges funding from U.K. EPSRC (EP/J018171/1, EP/N028694/1)
The authors thank the OSIRIS consortium at UCLA and IST for providing access to the OSIRIS 3.0 framework.

## Competing interests:

## Contributions

C.A. proposed the experiment, designed the experimental setup and performed the experiment, V.B.P. carried out PIC simulations, K.H.O., C.I.H., B.J.Y, T.G.P, P.K.S. and B.R.L. assisted the organization of the experiment, E.B. wrote the code used for data acquisition and processing, and J.H.S. and S.K.L. supported the laser operation and maintenance. C.A., H.T.K. and C.H.N. wrote the manuscript. All authors reviewed the manuscript.

# Supplementary material

| No shot | RMS Peak energy (MeV) | RMS Relative energy spread (%) | RMS Horizontal divergence (mrad) | RMS Vertical divergence (mrad) | Relative Charge (a.u) |
|---|---|---|---|---|---|
| 01 | 187 | 31.5 | 22.1 | 9.3 | 29.2 |
| 02 | 139 | 71.9 | 26.1 | 12.5 | 21.1 |
| 03 | 215 | 141.8 | 13.2 | 16.0 | 25.6 |
| 04 | 174 | 27.5 | 27.2 | 18.4 | 42.5 |
| 05 | 247 | 64.7 | 20.6 | 23.1 | 23.4 |
| 06 | 182 | 64.2 | 23.9 | 12.3 | 8.3 |
| 07 | 192 | 60.9 | 11.7 | 7.2 | 18.5 |
| 08 | 178 | 37.6 | 17.2 | 12.3 | 20.0 |
| 09 | 235 | 77.4 | 19.1 | 11.0 | 4.6 |
| 10 | 202 | 49.5 | 18.2 | 31.8 | 7.8 |
| 01n | 275 | 9.0 | 12.1 | 9.3 | 1.6 |
| 02n | 338 | 4.7 | 8.1 | 5.7 | 0.6 |
| 03n | 306 | 6.4 | 6.9 | 7.0 | 1.3 |
| 04n | 312 | 5.4 | 5.3 | 4.1 | 0.7 |
| 05n | 276 | 8.2 | 8.2 | 6.9 | 0.8 |
| 06n | 286 | 7.9 | 6.2 | 6.1 | 1.2 |
| 07n | 344 | 5.5 | 5.8 | 6.4 | 0.3 |
| 08n | 338 | 4.5 | 7.1 | 5.9 | 0.9 |
| 09n | 331 | 6.0 | 20.0 | 6.4 | 0.8 |
| 10n | 248 | 8.9 | 23.3 | 16.1 | 1.1 |

*Table 1: Summarized electron beam parameters for the selected shots shown in Figure 1 and Figure 2. The shot number with suffix "n" corresponds to the data taken in the presence of nanoparticles.*